# Models of Galaxies with Central Black Holes: Simulation Methods

Steinn Sigurdsson[1], Lars Hernquist[2] and Gerald D. Quinlan

*Board of Studies in Astronomy and Astrophysics, University of California, Santa Cruz, CA 95064*

## ABSTRACT

We present a method for simulating numerically the effect of the adiabatic growth of black holes on the structure of elliptical galaxies. Using a parallel self–consistent field code, we add black holes to N–body realizations of model distribution functions for spherical galaxies, using a continuous mass–spectrum. The variable particle mass, combined with a simple multiple timestep integration scheme, makes it possible to evolve the models for many dynamical times with $N \sim 10^6 - 10^8$, allowing high spatial and mass resolution. This paper discusses verification of the code using analytic models for spherical galaxies, comparing our numerical results of the effect of central black holes on the structure of the galaxies with previously published models. The intrinsic and projected properties of the final particle distribution, including higher order moments of the velocity distribution, permit comparison with observed characteristics of real galaxies, and constrain the masses of any central black holes present in those galaxies. Our technique is promising and is easily extended to axisymmetric and triaxial galaxies.

*Subject headings:* black holes — galaxies: nuclei — galaxies: structure — stars: stellar dynamics

---
[1] current address: Institute of Astronomy, Madingley Road, Cambridge, CB3 0HA, England
[2] Alfred P. Sloan Fellow, Presidential Faculty Fellow

## 1. Introduction

Theories of energy production from quasars and active galactic nuclei predict that many present–day galaxies contain central massive black holes (MBHs) with masses $M_{BH} \gtrsim 10^7 M_\odot$ (Rees 1990). Strong observational efforts have revealed many candidate galaxies suspected of harboring MBHs, but conclusive proof remains elusive (see reviews by Dressler 1989, Gerhard 1992, Kormendy 1992). In tandem with past observational studies, and proposed studies to be made with the refurbished Hubble Space Telescope, Keck and other ground based observatories (*eg.* Sargent et al., 1978, Dressler & Richstone 1990, Lauer et al., 1992a, Lauer et al., 1992b, Stiavelli et al., 1993, Crane et al., 1993, Harms et al., 1994, van der Marel et al., 1994, Kormendy et al., 1994), there has been intense theoretical effort to constrain the masses of these claimed central black holes and provide theoretical predictions of their consequences (*eg.* Bahcall & Wolf 1976, Young 1980, Duncan & Wheeler 1980, Binney & Mamon 1982, Tonry 1983, Richstone & Tremaine 1985, Shapiro 1985, Binney & Petit 1989, Lee & Goodman 1989).

Using N–body simulations and analytic techniques, it is possible to investigate the effects of a central MBH on the dynamics of stars in galaxies, and predict the range of observational properties of real galaxies containing MBHs. A variety of approaches have been used to model galaxies containing MBHs (*eg.* Young 1980, Norman et al., 1985, Richstone & Tremaine 1985), demonstrating that the observed structure of many galaxies is consistent with the presence of a MBH. At the same time, some authors have also shown that the observations may be accounted for by models of galaxies with no black holes (Duncan & Wheeler 1980, Binney & Mamon 1982; but see Merritt 1987).

When applied to this problem, most analytic techniques are restricted to spherical, or at best axisymmetric models of galaxies. Real galaxies are generally triaxial, and have small but measurable bulk rotation, which may strongly affect the influence a central MBH can have on the structure of the galaxy (Gerhard & Binney 1985, Pfenniger & de Zeeuw 1989). Using N–body realizations of galactic models, one can directly examine the consequences of triaxiality, investigate instabilities (Merritt 1987, Palmer & Papaloizou 1988), and analyze the orbital populations and observational signatures of MBHs. In the limit of large $N$, simulations approach the intrinsic "graininess" of real galaxies, where the luminosity is supplied by a finite number of effective point sources. Other authors have performed N–body simulations of the structure of galaxies with central MBHs, notably Norman, May & van Albada (1985) and Hasan & Norman (1990). As noted by Binney & Petit (1989), previous simulations (with $N \sim 10^4$) have been limited by poor resolution, spurious numerical relaxation or have employed unrealistic distribution functions.

Here we present a technique for simulating the adiabatic growth of MBHs in galaxies, and discuss its application to spherical models. We compare our results with earlier theoretical results, in which MBHs were assumed to grow adiabatically in some background stellar distribution, and the final distribution function was calculated assuming the action variables remained constant (Young 1980, Quinlan et al., 1994). The comparison serves to verify the code and analysis methods, validate the "adiabatic growth" approximation for adding the BH to the galaxy, and check the final system for orbital stability in the presence of a BH. The code permits straightforward orbit classification for subsets of particles, and the analysis of the evolution of orbit families as the potential evolves.

Using the Self-Consistent Field (SCF) method (Hernquist & Ostriker 1992), we can now follow large ($N \sim 10^6 - 10^8$) self-gravitating models for many dynamical times, allowing the central regions of the galaxies to be well-resolved, and making possible statistically significant assertions about the spatial gradients of observed properties of the models. The broadband light profiles of elliptical galaxies are dominated by post–main sequence stars, constituting $\sim 1\%$ of the total number of stars. Our particle resolution is approaching the discreteness of the light tracing the potential; the actual potential is likely smoother, whether the potential is dominated by stars or dark matter.

In this paper we examine how to generate and evolve large $N$–body realizations of a family of distribution functions to which central MBHs are added. Using massively parallel processing (MPP) systems, we can integrate realistic calculations for $\sim O(100)$ dynamical times, and analyze the intrinsic and projected properties of the models as the MBH forms. We investigate the stability of the models, the limits of the assumption of adiabatic growth, and the true spatial resolution of our realizations as a function of $N$ and $M_{BH}$. We develop and verify schemes for generating multi–mass realizations of our choice of distribution function, a multiple timestep scheme to allow fast integration despite the large range in intrinsic timescales across our models, and parallelized analysis routines to calculate the projected moments of the distribution.

## 2. Models and Methods

### 2.1. The Self–Consistent Field Method

The simulation method we use is based on that described by Hernquist & Ostriker (1992). The potential of the galaxy, $\Phi(r,\theta,\phi)$, is expanded in an orthonormal set of basis functions. By choosing a suitable biorthogonal basis set, the density, $\rho(r,\theta,\phi)$, is represented by a similar expansion; namely

$$\rho(r,\theta,\phi) = \sum_{nlm} A_{nlm} \rho_{nlm}(r,\theta,\phi) \qquad (2\text{-}1)$$

$$\Phi(r,\theta,\phi) = \sum_{nlm} A_{nlm} \Phi_{nlm}(r,\theta,\phi) \qquad (2\text{-}2)$$

where the $A_{nlm}$ are the expansion coefficients for the basis chosen and

$$\nabla^2 \Phi_{nlm}(r,\theta,\phi) = 4\pi \rho_{nlm}(r,\theta,\phi). \qquad (2\text{-}3)$$

We choose as our standard zeroth order basis function the Hernquist model (Hernquist 1990, Hernquist & Ostriker 1992), with units $G=1, M=1, a=1$ (note Quinlan et al., 1994 used $a=1/3$), defined by

$$\rho_{000}(r) = \frac{1}{2\pi r(1+r)^3} \qquad (2\text{-}4)$$

$$\Phi_{000}(r) = \frac{1}{1+r}. \qquad (2\text{-}5)$$

The particles determine the $A_{nlm}$ through a numerical integration over the density, and move under the potential gradient derived from the expansion by equation (2-2), providing a completely self–consistent scheme for particle interaction. Each of the particles describing the density distribution may have different masses. The expansion coefficients can be saved during the time evolution and provide a compact snapshot of the time evolution of the density distribution. Together with a sampling of the particle phase space, the $A_{nlm}$ provide a straightforward means for numerical classification of orbit families.

Using this basis set we can well–represent a number of standard potential–density pairs for spherical galaxies, including the truncated isothermal sphere and the family of $\gamma$–models (also known as $\eta$–models) (Dehnen 1993, Tremaine et al., 1994). The Hernquist (or $\gamma=1, \eta=2$) model is our canonical example for a spherical galaxy containing no central MBH. All our results are compared with this default model, and the derived analytic results.

The SCF algorithm is very parallelizable. An efficient parallel implementation of this technique has been implemented on some MPP architectures, specifically the CM–5 and T3D, with other implementations under development (Hillis & Boghosian 1993, Hernquist et al., 1994). Using parallel SCF codes, we can integrate models with $N \sim 10^6 - 10^8$ for $\gtrsim 100$ dynamical times, on machines like the NCSA 512 processor CM–5. With $2^{23}$ particles, typical of a full scale simulation, a file describing the complete particle distribution $(m_i, x_i, y_i, z_i, vx_i, vy_i, vz_i, \Phi_i)$, requires 512Mb of disk space. Test models, such as we describe here, generally employ 512,000 particles. A typical test run, integrating a spherical model for $\sim 40$ dynamical times, requires one or two hours using either 128 or 256 processors of the CM–5.

### 2.2. The models

We generate initial conditions from the distribution function, $f(E)$. The model is truncated at some radius, $r_t \gg 1$, and particle coordinates are chosen using an acceptance–rejection algorithm. For equal-mass models, in which $m = 1/N$, we have routines to generate realizations of various distribution functions, notable, $\gamma = 0$, $\gamma = 1$ (Hernquist) and $\gamma = 2$ (Jaffe) models (Hernquist 1990, Jaffe 1983). Given an initial realization, we grow a black hole at its center. We choose some black hole mass, $M_{BH}$, and a time to grow the black hole, $t_{BH}$, and then introduce a mass, $M(t)$, for $t \leq t_{BH}$,

$$M(t) = M_{BH}\left(3\left(\frac{t}{t_{BH}}\right)^2 - 2\left(\frac{t}{t_{BH}}\right)^3\right) \quad (2\text{-}6)$$

$$M(t) = M_{BH} \text{ for } t > t_{BH}, \qquad (2\text{-}7)$$

having an associated (softened) potential,

$$\Phi_{BH}(r,t) = M(t)/\sqrt{r^2 + \varepsilon_{BH}^2}. \qquad (2\text{-}8)$$

The particles move under the combined potential gradient, $\nabla(\Phi + \Phi_{BH})$. The choice of $\varepsilon_{BH}$ is driven by the spatial resolution in the center; higher $N$ requires smaller $\varepsilon_{BH}$. For the equal-mass tests $\varepsilon_{BH}$ ranges from $2.5 \times 10^{-3}$ to $10^{-2}$.

For a $\gamma$-model, the mass interior to some radius, $r$, is given by

$$M(r) = \left(\frac{r}{1+r}\right)^{3-\gamma}. \qquad (2\text{-}9)$$

In equal-mass models, the number of particles, interior to $r \ll 1$, $N_r$, is then $N_r \approx N \times r^{3-\gamma}$ (Dehnen 1993). For $\gamma = 1$, and $N = 10^6$, we have $N_{r=0.1} \approx 10^4$, and $N_{r=0.01} \approx 10^2$. So even with $N \sim$ few $\times 10^6$, the model has no statistical resolution for $r \lesssim 10^{-2}$, just where we expect unambiguous signatures of a central MBH (Young 1980, Quinlan et al., 1994).

To improve the resolution of our models we introduce a multi-mass scheme, where the mass of a particle, $m$, is a continuous variable. We rewrite the distribution function, $f(E) = N(E, J) \times m(E, J)$, where $J$ is the particle angular momentum. Then we define

$$r_p = \sqrt{\frac{(rv_t)^2}{E + M/a}} \qquad (2\text{-}10)$$

$$m(E, J) = m_n r_p^\lambda \text{ for } r_p \leq r_m, \qquad (2\text{-}11)$$

where $v_t$ is the transverse velocity of the particle, $M$ and $a(=1)$ are the total mass and scale radius, $r_m$ is some limiting radius ($= 1$ in practice), and $m_n$ is a normalizing scale factor. $r_p$ is an approximation to the particle's pericenter, usually accurate to within a factor of two. For $r_p > r_m$, $m = $ const. $\lambda$ controls the range of masses used.

In practice we calculate models for $\lambda = 0, 0.5, 0.75, 1.0$. $\lambda = 0.5$ provides a moderate mass range and $\lambda = 1.0$ provides a more extreme mass range. We suppress the mass variation outside $r_m$ so that the representation of the halo of the model remains tolerably smooth, and statistically robust. Figure 1 shows the mean particle mass as a function of $r$ for $\lambda = 0.5, 1.0$ in a $\gamma = 1$ model. The $\lambda = 1$ model provides an order of magnitude increase in mass range over $\lambda = 0.5$, and correspondingly larger numbers of particles at small radii. We tested multi-mass, $\gamma = 1$, models for spurious relaxation and mass segregation. No mass segregation was found, as might be expected by the nature of the force calculation, and any evolution was consistent with relaxation to virial equilibrium due to truncation of the initial conditions, and numerical relaxation due to the finite number of particles. The relaxation was not significant for the large numbers of particles we employ in our simulations.

With $\lambda = 1$, we gain over two orders of magnitude in particle resolution near the center over equal-mass models. In order to make use of the improved spatial resolution, we are forced to a smaller $\varepsilon_{BH} \lesssim 10^{-3}$, providing more than a factor of ten gain in spatial resolution. In the absence of a black hole, central velocities are $\sim 1$ in our units. With a central black hole, the velocity increases as $r^{-1/2}$ down to the smoothing length, requiring a correspondingly smaller timestep for the integration. For a small $\varepsilon_{BH} \lesssim 10^{-3}$, and a large $M_{BH}(\gtrsim 0.01)$, this forces a prohibitively large number of timesteps per dynamical time. To circumvent this problem, we implemented a simple, efficient multiple timestep scheme.

Ideally we want the particles near the black hole to move on smaller timesteps than particles at large radii. The particles requiring small timesteps constitute a near negligible fraction of the total mass, and the potential they are moving in is, in general, dominated by the black hole. The self-gravity of the rapidly moving particles is negligible. To retain parallelism we want to avoid treating a subset of the particles differently from the majority. We implemented a two level timestep scheme, whereby $\Phi_{BH}(r)$ is updated more often than $\Phi(r)$. We choose some controlling timestep, $dt$, sufficient for a precise integration of the self-gravitating particles. Every $dt$, we recalculate the expansion coefficients and the associated potential. In addition, we introduce a short timestep, $dt_s = dt/2^q$. We update $\Phi_{BH}(r)$ for each particle every short timestep, and integrate the particle motion using a simple leapfrog integrator, under the updated black hole potential, holding $\Phi(r)$ fixed. For $r \ll a$, where velocities are high, $\nabla \Phi \approx \nabla \Phi_{BH}$, and at larger radii, the potential changes slowly compared to $dt_s$. This scheme is very fast. $q = 10$ requires only a factor of 2 longer CPU time, while gaining a factor of 1024 in time resolution for motion near the center, and a corresponding factor of 100 increase in spatial resolution. Using this scheme, integrating 512,000 particles with *no* black hole present, and $q = 10$, energy is conserved to $\delta E/E \sim 10^{-6}$ over 40 dynamical times, implying that the scheme is robust.

### 2.3. Analyzing the data

To analyze our simulations, we concentrate on two types of output. The intrinsic properties of the distribution, such as volume density, dispersion, kurtosis and anisotropy, are useful for comparing with theoretical models. The projected properties of the model must also be determined for direct comparisons with observations. We derive both the surface density and classical velocity moments, and the Gauss–Hermite moments of the projected velocity distribution, following Gerhard (1993) and van der Marel & Franx (1993).

For comparison to theory, we bin a model into annular zones, adjusting the width of the zones to contain equal numbers of particles. Summing over the mass weighted particle distribution in each zone, we derive the density and surface density, true and projected dispersion, and the projected mean velocity, third and fourth moments of the velocity (skewness and kurtosis, $k = <v_p^4>/<v_p^2>^2$), as well as the anisotropy, $\beta = 1 - <v_t^2>/<2v_r^2>$, where $v_r$ is the radial velocity. This choice of binning allows constant statistical sampling across the model, maximizing the signal for models with symmetry, and provides direct comparison with theory, specifically to the results of Young (1980) and Quinlan et al. (1994). Typically we use $n_b = 2000$ particles per annular zone.

In practice, observers do not fold their data into constant light annuli. For comparison with observations, we sample our models with synthetic "slits" and apertures. The aperture sampling is simply done by considering all particles inside some projected radius $R_0$, and considering the projected properties of the distribution as $R_0$ varies.

For "slit" projection we consider the projected properties of a rectangular region, projected down the (arbitrary) z-axis of our model. The "slit" has some length, $y_s$, and width, $x_s$, divided into $n_s$ rectangular boxes along the y-axis. The slit is symmetric about the center along the y-axis but may be offset from the center along the x-axis by some value, $x_o$. For spherical models, slit analysis does not provide any additional information. The method was developed with consideration for future work where we will investigate axisymmetric and triaxial models.

For each box or annulus, we calculate a surface density, $\Sigma(r)$, projected mean velocity, $\bar{v}_z(r)$, projected dispersion, $\sigma(r)$, projected skewness and kurtosis. In addition, we calculate the Gauss–Hermite moments, $s_i(r)$ and $h_i(r)$ (Gerhard 1993, van der Marel & Franx 1993). Following Gerhard (1993), we define, $w_j(r) = (v_{zj} - \bar{v}_z(r))/\sigma(r)$, and

$$s_i(r) = \frac{1}{n}\beta_i \sum_{j=1}^{n} H_i(w_j) e^{-\frac{1}{2}w_j^2}, \qquad (2\text{-}12)$$

where the $H_i$ are the usual Hermite polynomials, and $\beta_i = 1/\sqrt{2^{i-1}i!}$ are normalizing constants. As noted by van der Marel et al. (1994), the observed velocity distribution is not fit for the true $\bar{v}_z$, $\sigma$, but rather a "best fit" Gaussian profile is derived from the line profile. Hence they derive a "best fit" Gauss–Hermite fit, $h_i(r)$, defined as for the $s_i$, but using "best fit" $\bar{v}'_z(r)$ and $\sigma'(r)$, such that $h_1(r) = 0 = h_2(r)$. We follow Heyl et al. (1994) and derive the $h_i(r)$ moment coefficients iteratively from the $s_i(r)$ moments. Given $\bar{v}_z, \sigma, s_1, s_2$, define, $\bar{v}'_{z1}(r) = \bar{v}_z(r)$, $\sigma'_1(r) = \sigma(r)$, and solve for $h_i(r)$, then define

$$\bar{v}'_{z(l+1)} = \bar{v}'_{z(l)} + h_1 \times \sigma'_l \qquad (2\text{-}13)$$
$$\sigma'_{l+1} = \sigma'_l + h_2 \times \sigma'_l, \qquad (2\text{-}14)$$

and solve for $h_1, h_2$ recursively until $h_{1,2}(r) \leq \delta$. In this paper we choose $\delta = 10^{-6}$, though the solution is not sensitive to the exact choice for $\delta$, in general. About a dozen iterations are required for $h_{1,2}$ to converge. The algorithm is easily parallelizable, and a version of the slit analysis has been implemented on the CM–5. Using logical parallel mask constructs on the phase space arrays, expensive sorts may be avoided and the data reduced rapidly. An $N = 8,388,608$ model is analyzed in less than 100 seconds using 256 nodes of the CM–5.

In the future, we also intend to generate synthetic line–profiles, using a blend of stellar line–profiles with appropriate offset, drawn from a library of model stellar lines. From such synthetic "observations" we can test how well analysis of actual observations can reproduce the underlying dynamics when convolved with seeing errors and observational noise.

## 3. Results

### 3.1. Tests of Parameters

Preliminary tests of our code used the truncated isothermal sphere as the basic model. Comparison was made with the classic paper by Young (1980) and our results agreed both with those of Young and our separate re-analysis (Quinlan et al., 1994).

In what follows our canonical test case is a 512,000 particle equal–mass realization of a Hernquist model. All numerical results are compared with the basic results derived from that model, and the corresponding analytic work by Quinlan et al. (1994). The model is truncated at $r_t = 300a$. The mass enclosed, $M(r) = r^2/(1+r)^2$ (= 0.993 for $r_t = 300$), is renormalized to unity by uniformly rescaling the particle masses. The resulting model is slightly sub-virial and settles into equilibrium in a few dynamical times. This transient evolution does not significantly affect the response to a growing central black hole, provided $r_t$ is sufficiently large. For $\gamma = 2$, the center of mass is significantly offset from the density maximum for $N \lesssim 10^6$, and the density cusp may be destroyed for $r \lesssim 10^{-2}$ if the black hole is not correctly centered on the density cusp. The displacement of the center of mass is also an issue with multi-mass models, where the outer particles are more sparsely sampled at fixed $N$.

We grow an MBH with mass $M_{BH} = 0.01$, smoothing length $\varepsilon_{BH} = 0.01$, over $t_{BH} = 20$ dynamical times (a dynamical time is defined naturally by $t = a/v$), using $dt = 2.5 \times 10^{-3}$. The model is allowed to settle for a further 20 dynamical times before the integration is terminated after 16,000 steps. For this reference spherical model, we use $n = 16, l = m = 0$. The properties of the initial and final distribution are shown in Figures 2 and 3. Figure 2 shows the characteristic Keplerian rise in dispersion due to the black hole. The cusp induced by the black hole is detectable for $r \lesssim 0.1$. At this spatial resolution and $M_{BH}$ there is no detectable anisotropy, which is consistent with the analytic predictions. Figure 3 shows $k - 3$ and $h_4$. The numbers are in agreement with the results of Quinlan et al. (1994). Note that contrary to the case of the truncated isothermal sphere, both $k - 3$ and $h_4$ are flat in the presence of a black hole at this resolution, and rise in the absence of a black hole, again in agreement with analytic results. The calculated value of both $k$ and $h_4$ at the smallest $R$ is limited by smoothing.

### 3.1.1. Integration Parameters

To test the robustness of our assumptions, we consider variations of the simulation parameters.

We first try using larger timesteps, $dt = 10^{-2}$, and find that the results agree with our canonical model, with an increase in $\beta$ at the center at $r = \varepsilon_{BH}$ that is not statistically significant. Increasing $dt$ by another factor of two leads to significant radial anisotropy due to integration errors as stars near the center are scattered by the black hole.

We did a run with $n = 32$, to check that the model was adequately resolved with the canonical $n = 16$. There was no significant difference between the $n = 16$ and $n = 32$ models, implying that $n = 16$ is sufficient. A smaller $n$ would be adequate for most of the models considered here; the $A_{nlm}$ suggest $n \approx 8$ would have sufficed for most of the runs, but for purposes of testing the code we chose to use $n = 16$. The variation of the kurtosis, $k - 3$, with $r$ was a little smoother with $n = 32$, showing more clearly the peak in $k$ near $r = 3 \times 10^{-3}$, but the difference was not statistically significant.

It is somewhat surprising that the shape of the MBH induced cusp is not sensitive to $n$ for $r \ll a$. We reproduce analytic estimates for the cusp to a smaller spatial resolution than are sampled by the relatively low order radial expansion. The reason this approach works, is that at these scales the potential gradient is dominated by the black hole, not the self-gravity of the responding change in the stellar distribution, which is negligibly small in comparison. Our tests show that the code does correctly reproduce the true dynamics of the problem, down to $r \sim 2\varepsilon_{BH}$.

To test the stability of the model, we ran a simulation with $l = 6, dt = 10^{-2}$. There was no significant growth of low $l, m$ modes; indeed for $n = 0, l = m = 1$ and $l = 1, m = 0$ the power in the modes decreased with time. There was large fractional variation in the coefficients for $n = 0, l = 5, 6, m = 1$, but the power was not significant due to numerical noise. There is some concern that Keplerian degeneracy of the fundamental modes of spherical systems supports slow modes that violate the adiabatic approximation (Tremaine [personal communication], Weinberg 1994), but this does not appear to be a problem for our models.

### 3.1.2. Model Parameters

We varied $t_{BH}$ to test the validity of the adiabatic growth approximation, using $t_{BH} = 20, 10, 1,$ and $5 \times 10^{-3}$. When $t_{BH}$ is too small, a large anisotropy is induced in the distribution as central particles are scattered by the rapidly changing potential. A significant radial anisotropy is observed with $t_{BH} = 1$, and for $t_{BH} = 5 \times 10^{-3}$ ($dt = 2.5 \times 10^{-3}$), the anisotropy is maximal ($\beta(r) \rightarrow 1$ for $r \rightarrow r_t$). However, $t_{BH} = 10$ shows no significant spurious anisotropy compared to the run with $t_{BH} = 20$, and we conclude that $t_{BH} \gtrsim 10$ is adequate for adiabatic growth approximation in spherical systems.

This result was independently verified by looking at the change in radial action of orbits in different (fixed) spherical potentials, to which a time varying

Keplerian potential was added. The radial action was conserved to within ~ 1%, provided the time scale for the Keplerian potential to grow was ~ 5 − 10 radial periods (Quinlan et al., 1994).

Finally, we grew smaller black holes, $M_{BH} = 10^{-3}$, with $t_{BH} = 10, dt = 10^{-2}, 2.5 \times 10^{-3}$, and $2.5 \times 10^{-4}$; for the last set, $\varepsilon_{BH} = 10^{-3}$. These models demonstrate the limitations of the equal–mass, single timestep models, even for $N \sim 10^6$, and should be compared with the multi–mass models presented below. As Figures 4 and 5 show, for $M_{BH} = 10^{-3}$ there is little observable signature of the MBH at this resolution. Even with 512,000 particles the signature of the MBH is barely noticeable. Decreasing $\varepsilon_{BH}$ allows the physics at small $r$ to be explored further, but the finite number of particles leads to a loss of signal, and the required $dt$ makes the simulation prohibitively expensive. As Figure 5 shows, $\varepsilon_{BH}$ is critical to estimating the kurtosis at the center of the model, while $h_4$ is a more robust estimator (cf. the behaviour of the short and long dashed lines). $k(0)$ is very sensitive to a few high speed particles near $r = 0$, which are poorly sampled with a finite $N$, and not present for feasible $\varepsilon_{BH}$ in the equal–mass, single timestep integration. $h_4$ is a more robust estimator, and shows the downturn at small radii expected from analytic calculations for this smoothing length.

We also ran equal–mass models for $\gamma = 0, 2$, verifying the results in Quinlan et al. (1994) to within the resolution of the models. The $\gamma = 0$ model shows the expected steep density cusp. For the $\gamma = 2$ model, our basis set cannot resolve the self-gravity of the central cusp well. Integrating a $\gamma = 2$ model with no central black hole $n = 16$ and $dt = 0.01$, the final state deviated significantly from the initial model only for $r \lesssim 0.03$ after 50 dynamical times. The MBH induced cusp for a $N = 512,000, \gamma = 2$, equal–mass model is limited less by the finite radial expansion than by $\varepsilon_{BH}$ and $N$. The steepening of the cusp due to the central MBH, seen in Quinlan et al., 1994, is observed to $r \gtrsim 2 \times \varepsilon_{BH}$, for $M_{BH} = 0.01$, and $\varepsilon_{BH} = 0.01$. It is important that the black hole be centered on the density cusp and not the center of gravity of the distribution. For $\gamma = 2$, even with 512,000 particles, the density maximum may be offset from the center of mass by $\gtrsim \varepsilon_{BH}$, and the evolution of the central density is to a flatter density profile, even with a central MBH added. The dispersion still shows an increase in the offset model, as it must for a locally virial distribution.

*3.1.3. Multi–mass Models*

Using a $\lambda = 0.5$ or $1.0$ multi–mass model provides a dramatic improvement in spatial and mass resolution. Figure 6 compares an $N = 512,000, \lambda = 0.5$ model with a equal–mass model. Figure 7 compares the resolution of $N = 512,000, \varepsilon_{BH} = 1 \times 10^{-3}$, $\lambda = 1.0$ and $0.5$ models. The anisotropy predicted by analytic models (Goodman & Binney 1984, Quinlan et al., 1994) is clearly evident in the multi–mass models for $r \gtrsim \varepsilon_{BH}$. For the $\lambda = 1.0$ model, the spatial resolution is less than the softening length, the number of particles at $r \lesssim \varepsilon_{BH}$ is large and statistically resolved on scales $< \varepsilon_{BH}$. The spatial resolution of the multi–mass model is a factor of 10 better than in the equal–mass case, and the dynamics can be followed to $\varepsilon_{BH} = 10^{-3}$ with the same number of timesteps using the multistep scheme. In order to obtain statistically significant slit "observations" of the central Gauss–Hermite moments, the multi–mass scheme is critical, as the number of particles in each slit box is much smaller than for the annular projection at fixed $N$ and $R$.

### 3.2. Velocity moments

While we evaluate both $s_i(r)$ and $h_i(r)$ (Gerhard 1993, van der Marel & Franx 1993), in practice the calculated moments are equivalent for the cases considered here. $s_4$ and $h_4$ may differ if $n_b$ is small and the realization of the average line profile is poorly sampled, or when the line profile becomes highly non–Gaussian, and the higher moments ($i \geq 6$) are large. The $h_i$ moments have the virtue that $h_1 = 0 = h_2$ by definition. By construction, $h_3 = 0$ for the models considered here, so all the information is contained in $h_4$ (higher moments may be calculated, but at this resolution are too noisy to be of use). $h_4$ is analogous to the kurtosis, but the exponential weighing suppresses the divergence of $k$ that makes it a poor estimator (van der Marel & Franx 1993). As a first approximation, $s_4 \approx h_4$, and $k - 3 \approx 8\sqrt{6}h_4$ for $h_4 \ll 0.03$.

As can be seen in Figure 8, $h_4(r)$ by itself is not an estimator for central MBHs. With the addition of the MBH, the $h_4$ of the Hernquist model approaches that of the isothermal sphere without a MBH. The value of $h_4(R = 0)$ is limited by $\varepsilon_{BH} = 10^{-3}$. For $R \gtrsim 2\varepsilon_{BH}$ the deviations of the velocity profile from a Gaussian are dominated by the intrinsic velocity profile of the underlying model. When averaging the velocity profile over an aperture centered on the MBH, the smoothing and particle resolution at small $R$ become very important. Both the smoothing of the potential, and the small number of particles leads to a deficiency in high velocity particles, which would lead to an increase in $h_4(0)$ if properly included.

Figure 9 explores the effect both of the smoothing and the particle resolution on the dispersion and $h_4$ measured in a circular aperture centered on the galaxy, as a function of the aperture size, $R_0$. $R_0$ may vary both because of improved instrument resolution for a particular galaxy, and from comparing galaxies at different distances. To mimic the true population of high velocity stars near the MBH we boost the particle velocities by a correction factor, $v'_z \mapsto v_z \times (\sqrt{R^2 + \varepsilon_{BH}^2}/R)^{1/2}$, and compare with the uncorrected profile. As the particles were integrated in the smoothed potential their spatial distribution inside $2\varepsilon_{BH}$ is not correct in any case. In particular, the integration scheme undersamples the pericenters of particles on orbits near $\varepsilon_{BH}$ biasing the velocity profile to lower velocities.

As can be seen in Figure 9, without the correction the best fit dispersion flattens at about $2\varepsilon_{BH}$. With the velocity correction the best fit dispersion continues to rise to the limit of the particle resolution. Over the range of apertures and models, the fit to a Gaussian line profile is surprisingly good, as can be seen from both $h_4$ and the ratio of best fit dispersion to true dispersion. The bottom left panel illustrates why the kurtosis is a poor estimator of the velocity profile, the central value is sensitive to both the aperture size, the model resolution and the velocity error induced by smoothing the potential. For $\lambda = 0.5$, and with no velocity correction, the central $h_4$ declines with $R_0$. With the velocity correction the decline in $h_4$ is at smaller $R_0$. With higher particle resolution, $h_4$ rises at the smallest $R_0$ well-sampled by the particles. As Figure 10 shows, this is entirely due to particles in the $\lambda = 1$ model reaching smaller radii and broadening the velocity profile. For $\lambda = 0.5$, even with the velocity correction, the number of particles inside $\varepsilon_{BH}$ is too small to raise $h_4$ significantly.

Figure 11 shows the slit analysis for a equal-mass $N = 8,388,608$ model and an $N = 512,000$, $\lambda = 0.5$ model. The multi-mass model is critical to retain particle resolution into the center of the model, while a large total $N$ is necessary to get statistically significant gradients for the moments of the distribution viewed through thin slits. At small $y$, the cusp in $\Sigma$ due to the presence of the black hole is clearly visible. The multi-mass model has relatively more particles at small $y$, and provides almost the same resolution in the central bin as the $8M$ model. The $512k$ multi-mass model becomes very noisy at $y \approx 0.1$, but tracks the larger model well at smaller $y$. This figure can be compared with Figures 4 and 5; the improved spatial resolution and higher particle number at small radii provides a clear signal of the surface density cusp and the change in dispersion and $h_4$ due to the MBH.

## 4. Conclusions

The work presented here is primarily to verify the technique developed, by comparing the results with analytic and numerical calculations performed using a completely different approach (Quinlan et al., 1994). Within the range of variables where our models are applicable, the results for intrinsic and projected properties of the models agree with those of our previous paper.

The models and analysis techniques are consistent with previous work, and other results in the literature (Young 1980, Goodman & Binney 1984, van der Marel 1994a,b). The basic algorithms for realization of the models, integration and analysis are correct.

Adiabatic growth is well approximated when black hole growth times of $\gtrsim 10$ dynamical times are used, in accordance with independent semi-analytic estimates of the variation of the action of particle orbits in time varying potentials. With $t_{BH}$ long enough we can be confident that we are seeing the true response of the stellar distribution. We also find no radial or low $l, m$ instabilities for these spherical models when a central black hole in the mass range $10^{-3} - 10^{-1}$ is added.

For spherical models, it is clear that the central surface density profile provides a poor diagnostic of the presence of a central MBH, while the projected dispersion provides a robust signal for the presence of an MBH, provided the observations have adequate spatial resolution. The refurbished Hubble Space Telescope is ideal for such observations of nearby galaxies.

Gauss-Hermite moments are a promising statistic for constraining anisotropic and non-spherical models with rising central dispersion, and may be used to discard extreme models producing rising central dispersion in the absence of a large central dark mass (eg. Duncan & Wheeler 1980). By themselves, the Gauss-Hermite moments do not provide a strong discrimi-nant for a central MBH. They are primarily useful for constraining bulk rotation, anisotropy, departures from symmetry and the distribution of the underlying model given the projected properties. With the Gauss–Hermite moments, particularly $h_3$ and $h_4$, we may preclude anisotropic models and triaxiality as the cause of rising projected central dispersion (Binney & Mamon 1982, Merritt 1987, van der Marel & Franx 1993). Measurements of $h_i$ at larger radii can help discriminate between model distribution functions that produce similar surface density profiles.

The $\gamma = 1$ model is a good approximation to the surface density profile of elliptical galaxies over a wide range of radii (Hernquist 1990). Setiing the scale length $a$ so that the effective radius, $R_e = 1.82a$ is similar to that for massive ellipticals ($\sim 3\,\text{kpc}$), provides a scale for the spatial resolution of our models. With a softening length of $10^{-3}a$ typical for our multi–mass models, we are resolving length scales of $\sim 1.5\,\text{pc}$. At a distance of $1\,\text{Mpc}$ this corresponds to an angular resolution of $\approx 0.3''$. For comparison, M87 is at $\approx 16\,\text{Mpc}$, and the HST with $0.1''$ resolution can explore spatial scales of $\sim 8\,\text{pc}$, corresponding to length scales of $\sim$ few $\times 10^{-3}$ in our models.

Physically, our model must break down in real galaxies for radii $r \ll 10^{-3}a$. Relaxation and stellar collisions become important at small radii in the presence of a cusp induced by a central black hole, so there is little point in exerting large computational effort to achieve much higher spatial resolution. It is still necessary to over–resolve the model. High velocity stars near the center contribute strongly to the velocity moments; and for triaxial models we expect particles on box orbits to explore the center of the model.

In real galaxies, we would expect MBHs to form at the density maximum, but it is possible that the BH may wander away from the center of the galaxy by an amount comparable to the spatial resolution used here, and that the real density cusp formed is also flattened. For example, black hole growth may occur during a merger with a satellite galaxy, and the nucleus of the satellite galaxy may remain a distinct subsystem at $r \gtrsim a$ while the central MBH grows by gas accretion. In that event, reflex motion of the MBH relative to the orbit of the satellite nucleus can cause the MBH to wander about the density maximum by distances of order $10\,\text{pc}$ on time scales of $O(10^8)\,\text{y}$. Our models indicate that in this case the final density cusp should be flatter than if the central MBH stays fixed at the density maximum, possibly flatter than the underlying density profile at the break radius.

Quinlan et al. (1994) found that the shape of the density cusp at the centers of galaxies is not a good diagnostic of the presence of a central MBH. Steep cusps (eg. $\gamma = 2$ models) may be present in the absence of a black hole, and the slope of a cusp induced by a central MBH may depend on the underlying stellar distribution even at constant stellar density. The results from our simulations of "off–center" MBH growth reinforces the conclusion that $\Sigma(R \to 0)$ is a poor indicator for the presence of an MBH, and that galaxies may have shallow or no resolved cusps even with a central MBH present.

It is perhaps surprising that the SCF method can be applied successfully to the problem of adiabatic growth of central masses in galaxies. The SCF provides a relatively coarse spatial resolution when the numbers of expansion terms used is small. In fact, we find excellent agreement between our simulation results and those obtained independently by analytic calculations, since the self–gravity of the response of the stellar background is negligible in the inner regions where the dynamics are dominated by the central mass. Out tests support the earlier claims by, eg. Hozumi & Hernquist 1994 and Johnston & Hernquist 1994, that the SCF technique will be a valuable tool for some problems in collisionless dynamics, and indicate that further theoretical studies, along lines similar to what we have presented here, are warranted.

The methods described here can be applied to large $N$ simulations of axisymmetric and triaxial models of galaxies. We hope to constrain self–consistent non–spherical models of galaxies containing central MBHs, and to survey the observable properties of families of different model galaxies containing MBHs. The method is also well suited for comparing particular, self–consistent realizations of theoretical models with the observed properties of individual galaxies.

We are grateful to Greg Bryan for help with parallelizing the SCF, and Jeremy Heyl for providing us with previously unpublished results. This work was supported in part by the National Center for Supercomputing Applications, the Alfred P. Sloan Foundation, NASA Grant NAGW–2422 and the NSF under grants AST 90–18526, ASC 93–18185 and the Presidential Faculty Fellows Program.

---

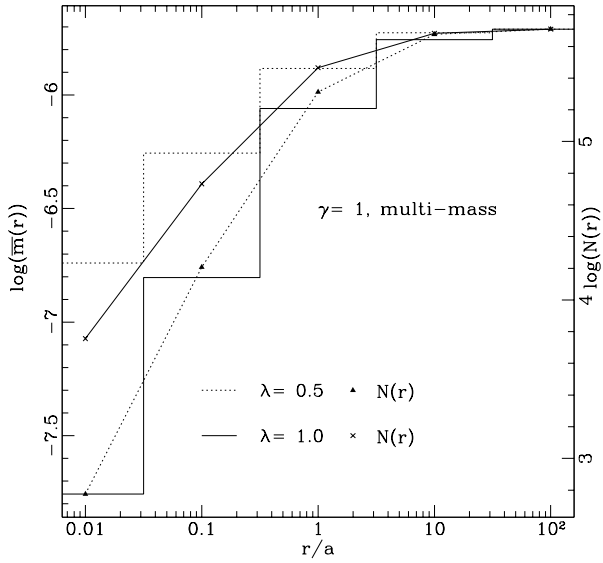

Fig. 1.— The mean mass, $\bar{m}(r)$, vs radius, $r$, for two variable mass $\gamma = 1$ models with $\lambda = 0.5$ and $1.0$. The plot shows a histogram of $\bar{m}$ for different radii, and curves giving cumulative particle number as a function of radius. The scale on the left y-axis refers to the histogramed data; the scale of the right y-axis refers to the connected points.

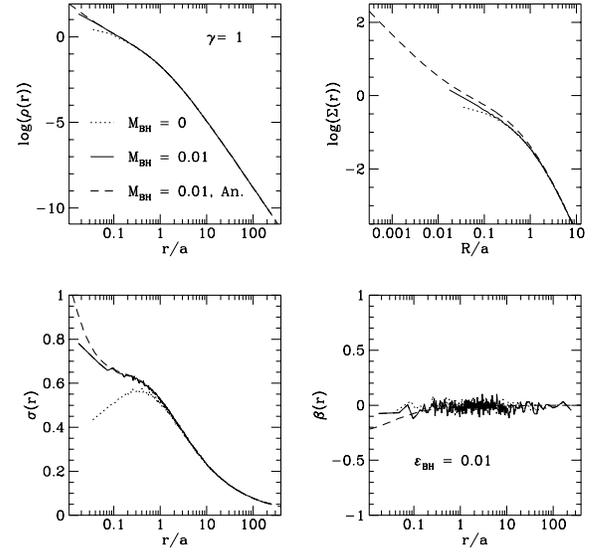

Fig. 2.— The properties of a 512,000 particle Hernquist model with no black hole and with a $M_{BH} = 0.01$ black hole, compared with the analytic results of Quinlan et al. (1994). Top left plot shows the volume density, with the cusp clearly seen at small $r$. The top right plot shows the surface density profile vs projected radius, $R$. The bottom left plot shows the dramatic rise in dispersion with the introduction of the black hole. The bottom right hand corner shows the anisotropy, $\beta$. The plots are done using annular projection with $n_b = 2000$ particles per annular zone unless otherwise stated.

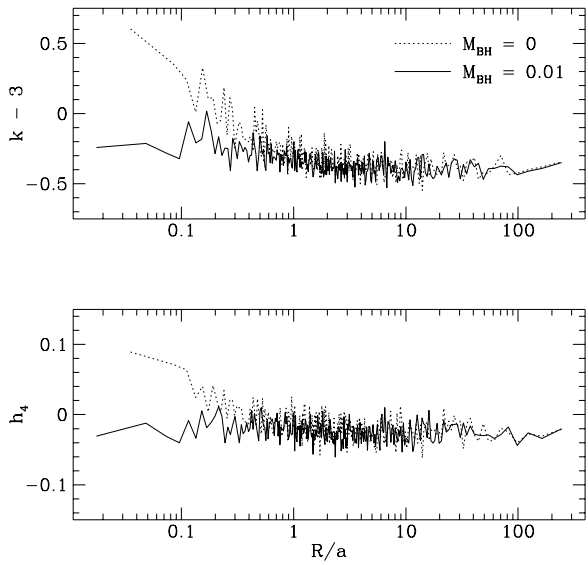
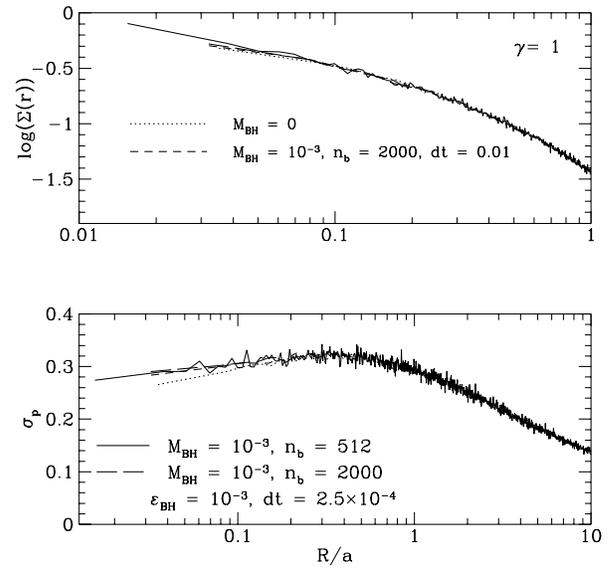

Fig. 3.— The projected kurtosis (minus three) and $h_4$ vs projected radius $R$ for our canonical equal-mass $\gamma = 1$ model with no central black hole (dotted lines) and with a $M_{BH} = 0.01$ central black hole (solid lines). The numbers shown in this figure were obtained using the annular projection discussed in the text.

Fig. 4.— The surface density and projected dispersion for a $\gamma = 1$ model, with a central black hole of mass $M_{BH} = 10^{-3}$ compared with the initial model, at three different resolutions. The short dashed line shows the model integrated with $\varepsilon_{BH} = 10^{-2}$, $dt = 10^{-2}$, the solid and long dashed lines show the same model with $\varepsilon_{BH} = 10^{-3}, dt = 2.5 \times 10^{-4}$, with different particle numbers per annulus.

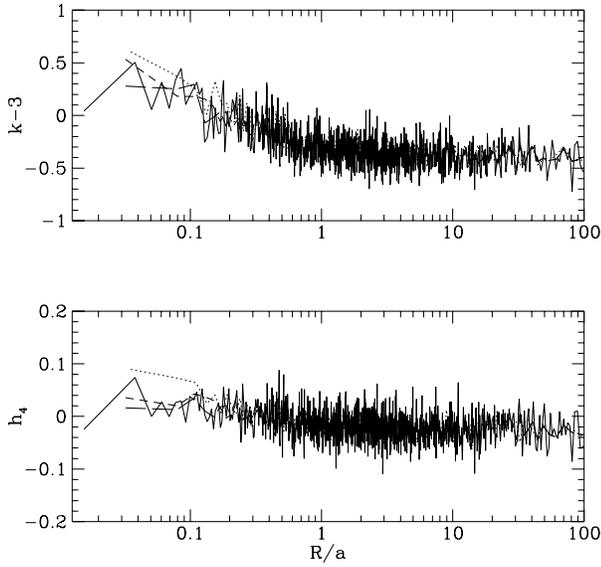
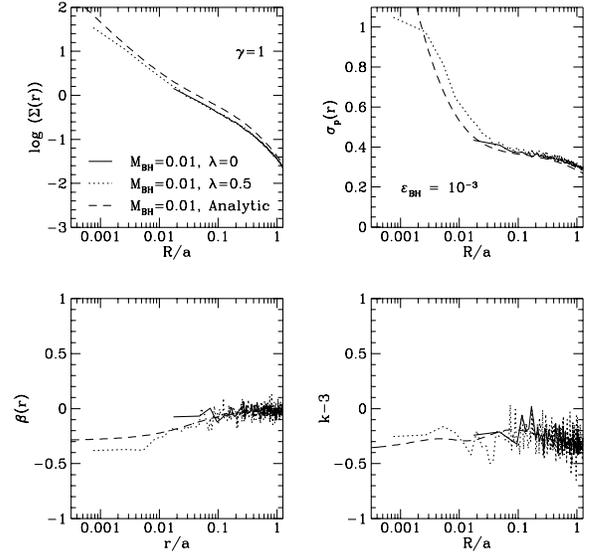

Fig. 5.— As for Figures 4 but showing $k-3$ and $h_4$. Compare the behaviour of the short dashed curve and the long dashed curve for $k-3$ and $h_4$. For $k-3$ the smoothed, long $dt$ integration approximates the no-MBH model.

Fig. 6.— The surface density, projected dispersion, anisotropy and projected kurtosis for equal-mass and $\lambda = 0.5$ multi-mass Hernquist models. The plots show the dramatic improvement in resolution with the multi-mass scheme. The slight offset of the (dashed) analytic curves is due to the finite width of the annular zones. The numerical calculations are centered on the mean radius of the zone, and the weighing of the averaged properties toward smaller radii within each zone leads to the offset observed compared to the analytically calculated properties.

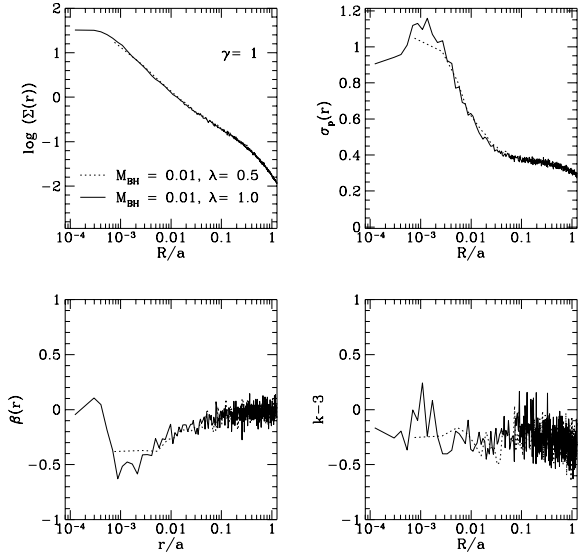
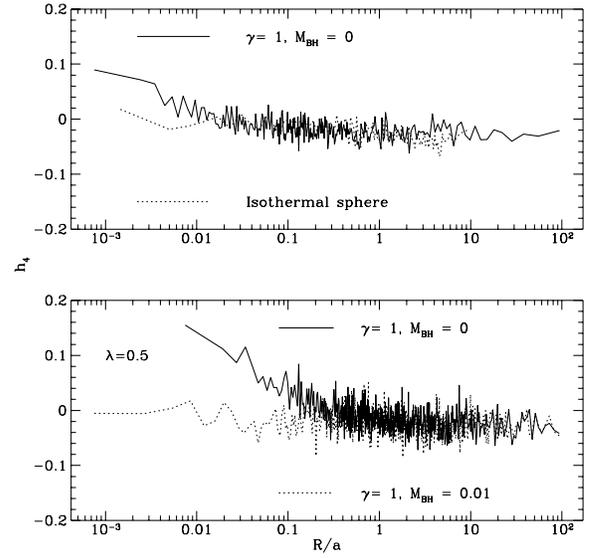

Fig. 7.— As for Figure 6, but comparing $\lambda = 1.0$ and $\lambda = 0.5$. At $\varepsilon_{BH} = 10^{-3}$, the $\lambda = 1.0$ over-resolves the central region. For non-spherical models, where the projection has to be done using "slits", the additional resolution is critical.

Fig. 8.— The fourth Gauss–Hermite moment for the isothermal sphere and Hernquist model with and without a central MBH. The upper panel shows $h_4$ for a equal-mass $N = 512,000$, $\gamma = 1$ model and a $N = 388,660$ equal-mass isothermal sphere. The radius of the isothermal sphere has been scaled down by a factor of 10 to match the Hernquist model scale. The lower panel shows $h_4$ for a multi-mass, $\lambda = 0.5$, $N = 512,000$, $\gamma = 1$ model.

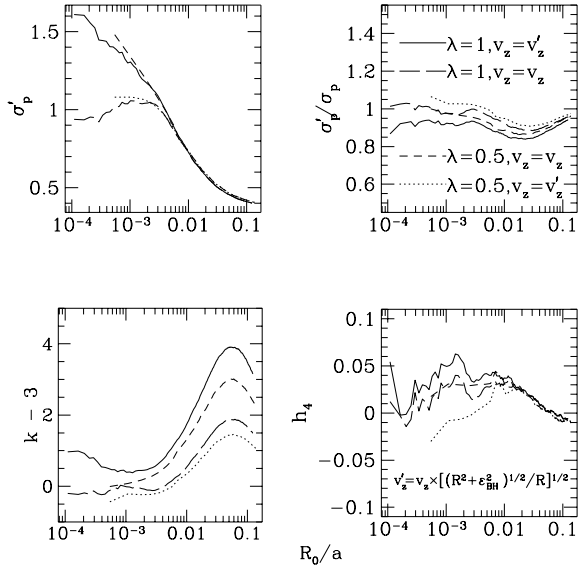
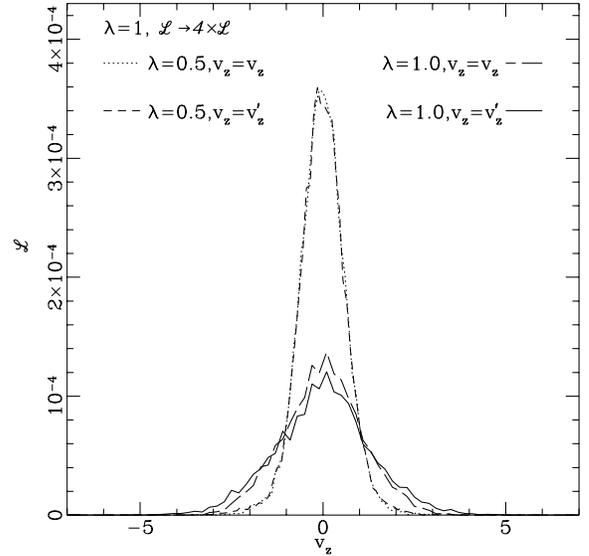

Fig. 9.— The effects of smoothing and particle number on the line profile. The plots show variable aperture fits to the center of the galaxies, as a function of aperture radius, $R_0$. The innermost point is for the innermost 512 particles, the aperture varies in increments of 512 particles. The top left panel shows the best fit dispersion, $\sigma'_p$ for four cases. The dotted line shows the $\lambda = 0.5$ model, the long dashed line shows the $\lambda = 1.0$ model. The short dash and solid lines show the same models, but with the velocity boosted to its Keplerian value to compensate for smoothing. The top right panel shows the ratio of the best fit dispersion to the true dispersion. The bottom right panel shows the variation in $h_4$. The bottom left panel shows $k - 3$ as a function of aperture size.

Fig. 10.— The line profile from the innermost projected 10,000 particles, weighted by particle mass. The line profile strength, $\mathcal{L}$, is shown with arbitrary normalization, with the $\mathcal{L}(\lambda = 1)$ scaled by a factor of four to fit on the plot.

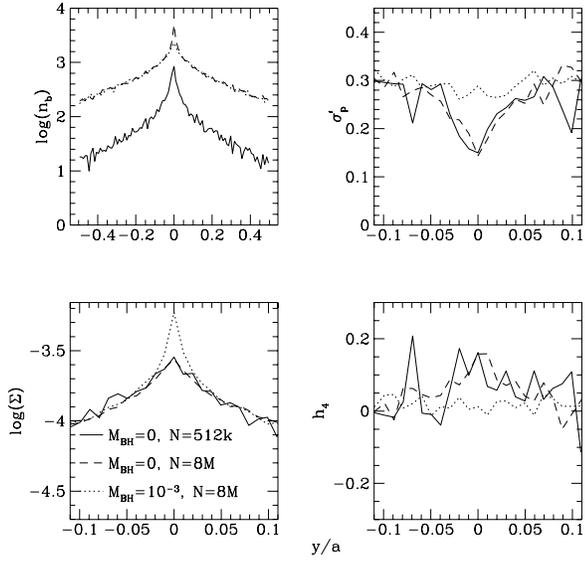

Fig. 11.— Slit projection of a equal–mass model with $N = 2^{23} = 8,388,608$ particles, with no central black hole and a black hole of mass $M_{BH} = 10^{-3}$, compared with the slit resolution of a $\lambda = 0.5$, $N = 512,000$ multi–mass model. The top left panel shows the number of particles per bin, as a function of the slit position, $y$. The slit width was $x_s = 0.01a$. The bottom left panel shows the resulting surface density (with arbitrary normalization). The top right panel shows the projected "best fit" dispersion, $\sigma'_p$ as a function of $y$. The bottom right panel shows $h_4(y)$, showing the importance of large $N$ to get reliable spatial gradients of projected moments with high resolution slit projections.

17